\documentclass[10pt]{article}
\pdfoutput=1
\usepackage{amssymb,amsmath,mathrsfs,enumerate}
\usepackage{graphicx,rotate,multicol}
\usepackage[margin=10pt,labelfont=bf]{caption}
\usepackage{cite}

\usepackage{color}
\usepackage{tikz,braket}
\long\def\rpl#1!!#2!!{\textcolor{red}{#1} \textcolor{blue}{#2}}

\let\tilde=\widetilde

\let\bar=\overline

\def \order(#1){{\mathcal O} \left(#1 \right)}

\evensidemargin=\oddsidemargin
\def\Eqn#1{Eq.\ (\ref{#1})}
\def\Eqs#1#2{Eqs.\ (\ref{#1}) and (\ref{#2})}
\allowdisplaybreaks

\title	{\LARGE \bf $S_3$ symmetry and the quark mixing matrix} 

\author {\sf Dipankar Das,$^{a,}$\footnote{dipankar.das@uv.es} \hspace{4pt}  Ujjal
  Kumar Dey,$^{b,}$\footnote{ujjaldey@prl.res.in} \hspace{4pt} Palash
B. Pal$^{c,}$\footnote{pbpal@theory.saha.ernet.in } \\[10pt]
\small\em $^a$ Departament de F\'{i}sica T\`{e}orica, Universitat de
Val\`{e}ncia and IFIC, Universitat de Val\`{e}ncia-CSIC� \\ 
\small\em Dr. Moliner 50, E-46100 Burjassot (Val\`{e}ncia), Spain\\ 
\small\em $^b$Theoretical Physics Division, 
		Physical Research Laboratory,
		Navrangpura, Ahmedabad 380009, India\\
\small\em $^c$Theory Division, Saha Institute of Nuclear
		Physics, 1/AF Bidhan Nagar, Kolkata 700064, India
}

\date{}

\begin{document}

\maketitle	

\begin{abstract}
We impose an $S_3$ symmetry on the quark fields under which two of
three quarks transform like a doublet and the remaining one as
singlet, and use a scalar sector with the same structure of $SU(2)$
doublets.  After gauge symmetry breaking, a $\mathbb{Z}_2$ subgroup of
the $S_3$ remains unbroken.  We show that this unbroken subgroup can
explain the approximate block structure of the CKM matrix.  By
allowing soft breaking of the $S_3$ symmetry in the scalar sector, we
show that one can generate the small elements, of quadratic or higher
order in the Wolfenstein parametrization of the CKM matrix.  We also
predict the existence of exotic new scalars, with unconventional decay
properties, which can be used to test our model experimentally.
\end{abstract}

\bigskip

Because of the discovery of a boson of mass about 126\,GeV at the
Large Hadron Collider (LHC) \cite{Aad:2012tfa, Chatrchyan:2012ufa}, we
are convinced of existence of elementary bosons of spin not equal to
1.  The discovery opens up the question whether there are more
particles of the same kind.  It is of course an experimental question,
but the possibility can be made attractive from a theoretical
standpoint if one can show that additional particles can help us
understand some properties or relate different parameters of the
Standard Model (SM).  This paper is an attempt in that direction.  We
show that the structure of the Cabibbo-Kobayashi-Maskawa (CKM) matrix
can be understood through an extended scalar sector and the breaking
of a discrete symmetry.

A very useful parametrization of the CKM matrix was given by
Wolfenstein \cite{Wolfenstein:1983yz}, which shows a hierarchical
pattern of the elements of the matrix.  We will show that this pattern
can be related, through an extended Higgs sector, to the breaking of a
discrete $S_3$ symmetry that we impose on the quartic terms of the
Lagrangian.  Admittedly, there have been many attempts to explain the
quark sector using $S_3$ symmetry\cite{Mondragon:1999jt,Kubo:2003iw,
  Chen:2004rr, Teshima:2005bk, Teshima:2012cg, Canales:2012ix,
  Canales:2013cga,Ma:2013zca}. But in most of them \cite{Kubo:2003iw,
  Chen:2004rr, Teshima:2012cg,Ma:2013zca} convenient relations among
the VEVs of different scalar multiplets have been assumed along with the
assumption that the scalar potential can produce such relations. On
the contrary, in a previous paper\cite{Das:2014fea} we studied in
detail the {\em most general} $S_3$ symmetric scalar potential with
three Higgs doublets and found that a $\mathbb{Z}_2$ subgroup of the
$S_3$ remains intact after the spontaneous symmetry breaking. In this
paper, we use this remnant $\mathbb{Z}_2$ symmetry to explain the mass
and mixing patterns in the quark sector.

The discrete symmetry group $S_3$ has two 1-dimensional and one
2-dimensional irreducible representations, which we will denote by
${\bf 1}$, ${\bf 1'}$ and ${\bf 2}$.  We pick a basis such that the
generators of the $S_{3}$ group in the ${\bf 2}$ representation is
given by,
\begin{eqnarray}
a = \begin{bmatrix} -\frac12 & \frac{\surd3}2 \\
- \frac{\surd3}2 & -\frac12
\end{bmatrix} \,, 
\qquad
b = \begin{bmatrix}\frac12 &
  \frac{\surd3}2  \\ \frac{\surd3}2 & - \frac12 
  \end{bmatrix}\,.
\label{gen}
\end{eqnarray}
Note that $a$ is of order 3, whereas $b$ is of order 2. The rest of
the elements can be obtained by taking products of powers of these two
elements. In this basis the quark fields are assigned the following
representations of $S_3$:
\begin{subequations}
\begin{eqnarray}
{\bf 2}   & : & \; \begin{bmatrix}Q_1 \\ Q_2 \end{bmatrix},
\; \begin{bmatrix}u_{1R} \\ u_{2R} \end{bmatrix},
\; \begin{bmatrix}d_{1R} \\ d_{2R} \end{bmatrix}\,, \\ 
 {\bf 1} & : & \; Q_3, \; u_{3R}, \; d_{3R} \,.
\end{eqnarray}
\end{subequations}
where the $Q_i$'s are the usual left-handed $SU(2)$ quark doublets,
whereas the $u_{iR}$'s and $d_{iR}$'s are the right-handed up-type and
down-type quark fields which are singlets of the $SU(2)$ part of the
gauge symmetry. Note that the square brackets, as in \Eqn{gen} as
well, denote the doublet representation of $S_3$, and has nothing to
do with the representation of the enclosed fields under $SU(2)$.
Similarly, in the Higgs sector, there are three $SU(2)$ doublets
$\phi_i$ ($i=1,2,3)$, and their transformation under the $S_3$
symmetry is as follows:
\begin{eqnarray}
{\bf 2} : \begin{bmatrix} \phi_1 \\ \phi_2 \end{bmatrix} \equiv \Phi
\,, \qquad {\bf 1}: \phi_3 \,. 
\end{eqnarray}
The most general scalar potential obeying gauge symmetry and $S_3$
symmetry has been given by many authors \cite{Pakvasa:1977in,
  Kubo:2004ps, Koide:2005ep, Teshima:2012cg, Machado:2012ed,
  Barradas-Guevara:2014yoa, Das:2014fea, Barradas-Guevara:2015rea},
and there is no need to repeat the expression here.  If we assume that
all scalar couplings allowed by the aforesaid symmetries are non-zero,
that all VEVs are real in order to avoid CP violation in the scalar
potential, then the minimization of the potential yields the
relation~\cite{Das:2014fea}
\begin{eqnarray}
v_1 = \surd{3} v_2 
\label{v1/v2}
\end{eqnarray}
assuming that the values of the parameters in the potential are such
that this minimum is favored over another one which has
$v_3=0$.  Of
course, $v=\sqrt{v_1^2+v_2^2+v_3^2}=246$~GeV appears in the masses of
the $W$ and $Z$ bosons. With the VEV relation in \Eqn{v1/v2}, one can
obtain an {\em alignment limit}\cite{Das:2014fea} where one of the
CP-even Higgs bosons, $h$, will have SM-like tree-level couplings with
the SM particles and therefore the LHC Higgs data can be explained.
The important consequence of \Eqn{v1/v2} is that a $\mathbb{Z}_2$
symmetry survives the spontaneous symmetry breaking, a symmetry that
is generated by the element which inflicts the transformation
\begin{eqnarray}
\Phi \to b\Phi
\label{Z2trans}
\end{eqnarray}
with $b$ defined in \Eqn{gen}, since this transformation leaves the
VEVs unaffected.  Note that the VEV relation of \Eqn{v1/v2} depends on
our choice of the basis for the doublet representation of $S_3$. In
another equivalent doublet representation of $S_3$, the relation
between the VEVs will change and the elements of $a$ and $b$ in
\Eqn{gen} will also change accordingly, but the vacuum will still
remain invariant under the transformation of \Eqn{Z2trans}.  In other
words, the existence of a remnant $\mathbb{Z}_2$ symmetry does not
depend on the choice of basis, and it is this fact that contains the
essential physics, as we will see shortly.\footnote{We note that the
  choice $v_3=0$ 
  with $v_2 = \surd3 v_1$ can lead to interesting dark matter candidates.}

We now present the most general Yukawa couplings involving the $u_R$
quarks that is consistent with the gauge and $S_3$ symmetries.
\begin{eqnarray}
\mathscr L_Y^{(u)} &=&
\null - y_1^u \Big( \bar Q_{1} \tilde\phi_3 u_{1R} 
+ \bar Q_2 \tilde\phi_3 u_{2R} \Big)
- y_2^u \Big\{ \Big( \bar Q_{1}\tilde\phi_2 + \bar
Q_2\tilde\phi_{1}\Big) u_{1R} + 
\Big( \bar Q_{1}\tilde\phi_{1} -
\bar Q_2\tilde\phi_2 \Big)u_{2R} \Big\} \nonumber \\*
&& \null - y_3^u \bar Q_3\tilde\phi_3u_{3R}
-y_4^u \bar Q_3 \Big( \tilde\phi_1 u_{1R} + \tilde\phi_2u_{2R} \Big)
-y_5^u \Big( 
\bar Q_1 \tilde\phi_1 + \bar Q_2\tilde\phi_2 \Big)
u_{3R} + {\rm h.c.}
\label{uYuk}
\end{eqnarray}
We have used the convention in which the lower component of the
$SU(2)$ doublets of Higgs multiplets are uncharged, and used the
standard abbreviation $\tilde\phi_i = i \sigma_2 \phi_i^*$.  The Yukawa
couplings of the $d_R$ quarks can be obtained by replacing $u_{iR}$ by
$d_{iR}$, $y_i^u$ by $y_i^d$, and $\tilde\phi_i$ by $\phi_i$ in
\Eqn{uYuk}.  The Yukawa couplings are in general complex, which can be
responsible for $CP$ violation.  It should be noted that the fields
$u_i$ and $d_i$ presented here do not represent physical quark fields.
Their superpositions which are eigenstates will be given later.

After symmetry breaking, the mass matrix that arises in the up-type
quark sector is the following:
\begin{eqnarray}
{\cal M}_u = \begin{pmatrix}
y_1^u v_3 + y_2^u v_2 & y_2^u v_1 & y_5^u v_1 \\
y_2^u v_1  & y_1^u v_3 - y_2^u v_2 & y_5^u v_2 \\
y_4^u v_1  & y_4^u v_2 & y_3^u v_3 \\
\end{pmatrix} \,.
\end{eqnarray}
This matrix can be easily block-diagonalized.  Taking
\begin{eqnarray}
X = \begin{pmatrix}
\frac12 & -{\surd3 \over 2} & 0 \\
{\surd3 \over 2} & \frac12 & 0 \\
0 & 0 & 1 
\end{pmatrix} \,,
\end{eqnarray}
we find that
\begin{eqnarray}
{\cal M}_u^{\rm block} \equiv X {\cal M}_u X^\dagger = 
\begin{pmatrix}
y_1^u v_3 - 2 y_2^u v_2 & 0 & 0 \\
0 & y_1^u v_3 + 2 y_2^u v_2 & 2 y_5^u v_2 \\
0 & 2 y_4^u v_2 & y_3^u v_3  \\
\end{pmatrix} \,,
\label{Mblock}
\end{eqnarray}
The mass matrix for the down-type quark is obtained by replacing all
$y^u$'s by the corresponding $y^d$'s.  It can also be
block-diagonalized using the same matrix $X$.

We can now identify the singleton blocks of the mass matrices to be
the masses of the third generation of quarks.  For example, the
$t$-quark mass will be given by
\begin{eqnarray}
m_t = \Big| y_1^u v_3 - 2 y_2^u v_2 \Big| \,,
\label{mt}
\end{eqnarray}
and a similar equation for $m_b$.  In other words, starting from the
original basis of quark fields that we had denoted by $u_1,\, u_2,\,
u_3,$ we have reached a new basis defined by
\begin{eqnarray}
\begin{pmatrix}t \\ c' \\ u' \end{pmatrix} = 
X \begin{pmatrix}u_1 \\ u_2 \\ u_3 \end{pmatrix} \,.
\label{newbasis}
\end{eqnarray}
The reason for the block-diagonal nature of the matrix in \Eqn{Mblock}
can be understood very easily from this new basis.  Notice that
\Eqn{newbasis} implies that 
\begin{eqnarray}
t = \frac12 (u_1 - \surd{3} u_2) \,.
\end{eqnarray}
It can be easily checked, using the $\mathbb{Z}_2$ generator, $b$,
that appears in \Eqn{gen}, that this combination changes sign under
the remnant $\mathbb{Z}_2$ transformation, i.e., it is
$\mathbb{Z}_2$-odd.  The other two members in the new basis, $c'$ and
$u'$, are $\mathbb{Z}_2$-even.  Because the $\mathbb{Z}_2$ symmetry
remains intact, there is no mixing between states which are odd under
it with states which are even.

This block structure has a very important consequence on the CKM
matrix, which is the main point of our article.  In order to obtain
the physical eigenstates, we still need to further rotate the
$2\times2$ block that remains in ${\cal M}^{\rm block}$.  In other
words, we can find a bi-unitary transformation such that
\begin{eqnarray}
U_L^\dagger{\cal M}_u^{\rm block}U_R   = {\cal
  M}_u^{\rm diag} = \mathop{\rm diag} (m_t, m_c, m_u) \,.
\label{Mdiag}
\end{eqnarray}
Both $U_L$ and $U_R$ would be block-diagonal.  We can take $U_L$ to be
of the form
\begin{eqnarray}
U_L = \begin{pmatrix}
1 & 0 & 0 \cr
0 & \cos \theta_u & -\sin\theta_u \\
0 & \sin\theta_u & \cos\theta_u 
\end{pmatrix} \,,
\label{UL}
\end{eqnarray}
with the understanding that all phases can be absorbed in $U_R$.
Therefore, combining \Eqs{Mblock}{Mdiag}, we obtain that
\begin{eqnarray}
{\cal M}_u^{\rm diag} ={\cal U}_L^\dagger {\cal M}_u {\cal U}_R  \,,
\label{Mudiag}
\end{eqnarray}
where
\begin{eqnarray}
{\cal U}_L =  X^\dagger U_L  \,.
\end{eqnarray}
The relation between original states $d_{iL}$ ($i=1,2,3$) and the mass
eigenstates in the down sector will have a similar form, governed by
the matrix
\begin{eqnarray}
{\cal D}_L =  X^\dagger D_L  \,,
\end{eqnarray}
where $D_L$ is a matrix like $U_L$, except with a different angle
$\theta_d$.  The CKM matrix is then given by\footnote{The first column
  and the first row of the CKM matrix correspond to the $b$ and the
  $t$ quarks merely because we do not want to disturb the notation of
  Ref.~\cite{Das:2014fea}.  We could have easily
  taken $u_2$ and $u_3$ as part of an $S_3$ doublet in Eq (2), and
  then we could have put their $\mathbb Z_2$-odd combintation in the
  third row.}
\begin{eqnarray}\label{Z2CKM}
V_{\rm CKM} =  {\cal U}_L^\dagger {\cal D}_L = \bordermatrix{\text{ }
  & \mathbf{b} & \mathbf{s} & \mathbf{d} \cr 
               \mathbf{t} &1 & 0   & 0 \cr
               \mathbf{c} & 0  & \cos\theta_{C} & -\sin\theta_{C} \cr
               \mathbf{u} & 0 & \sin\theta_{C} & \cos\theta_{C}
}
 \,,
\end{eqnarray}
where $\theta_C =  \theta_d - \theta_u$.

There are some interesting points to note here.  First, the CKM matrix
does not depend on the matrix $X$ that was used only to define an
intermediate basis to understand the effect of the remnant
$\mathbb{Z}_2$ symmetry.  Second, with the $\mathbb{Z}_2$ symmetry
intact, the CKM matrix is block diagonal.  This is expected, since the
$W$-boson, being $\mathbb{Z}_2$-even \cite{Das:2014fea}, cannot couple
a $\mathbb{Z}_2$-even quark to another which is $\mathbb{Z}_2$-odd.
This conclusion, i.e., the zeros of the CKM matrix that appears in
\Eqn{Z2CKM}, will not be modified by loop corrections because of the
unbroken $\mathbb{Z}_2$ symmetry.

In order to obtain the realistic CKM matrix, one therefore has to
break the $S_3$ symmetry in the Lagrangian itself.  We assume that
there are soft terms in the scalar potential which are not $S_3$
symmetric.  For example, we can consider
\begin{eqnarray}
V_{\rm soft} = \mu_{13}^2 (\phi_1^\dagger \phi_3 + \phi_3^\dagger
\phi_1) \,,
\label{soft}
\end{eqnarray}
with $\mu_{13}^2$ being much smaller than the other bilinear
parameters.  The presence of this term will slightly modify the VEV
relation of \Eqn{v1/v2}.  Let us denote the changed relation by
\begin{eqnarray}
v_1 = \surd{3} v_2 + \Delta \,,
\label{delta}
\end{eqnarray}
where $\Delta \ll v_2$.  This deviation $\Delta$ from \Eqn{v1/v2} is
what is important, and not the details of the soft-breaking terms,
since the generic form of \Eqn{delta} is obtained even if we make some
other choices in \Eqn{soft}.  The modified mass matrix in the up
sector is therefore given by
\begin{eqnarray}
\tilde {\cal M}_u = {\cal M}_u + \begin{pmatrix}
0 & y_2^u \Delta & y_5^u \Delta \\
y_2^u \Delta & 0 & 0 \\
y_4^u \Delta & 0 & 0 \\
\end{pmatrix} \,.
\label{Mupert}
\end{eqnarray}
We now need to modify the diagonalizing matrices.  Instead of the
prescription of \Eqn{Mudiag}, we will now need to use
\begin{eqnarray}
{\cal M}_u^{\rm diag} = {\mathscr
  U}_L^\dagger \tilde{\cal M}_u {\mathscr U}_R\,, 
\end{eqnarray}
where we can define ${\mathscr U}_L$ (similarly for ${\mathscr U}_R$
also) in the form
\begin{eqnarray}
{\mathscr U}_L = {\cal U}_L {\mathfrak U}_L \,,
\label{Ucorr}
\end{eqnarray}
where ${\mathfrak U}_L$ is close to the unit matrix which takes into
account the effect of very small $\Delta$ in \Eqn{delta}.  The task
now is to determine the form of this matrix.

For this, let us look back at \Eqn{Mblock}, in particular at the first
two diagonal elements of the matrix.  The top quark mass is given in
\Eqn{mt}, which has to be large.  One can ask which of the two terms
dominates in the expression.  If any one term is considerably larger
in absolute value than the other term, the 22-element of the matrix
${\cal M}_u^{\rm block}$ will roughly be equal to the 11-element.
That would be a disaster, because the trace of the lower $2\times2$
block, which should be of the order of charm quark mass, would be then
close to $m_t$ in absolute value.  The only way this problem can be
avoided, i.e., the 22-element remains much smaller than the
11-element, is by having both terms almost equal in magnitude, so that
their magnitudes add up in $m_t$ but largely cancel in the 22-element.
This means that, with comparable VEVs, both $y_1^u$ and $y_2^u$ will
have to be larger than the other Yukawa couplings by about an order of
magnitude.  Thus the 12 and 21 elements of the correction matrix of
\Eqn{Mupert} are much larger than the 13 and 31 elements.  So we
anticipate a form for the correction matrix ${\mathfrak U}_L$ where the
13 and 31 elements will be down by a power of some small (but not very
small) parameter $\lambda$, which will be specified shortly.

There will be a similar correction matrix ${\mathfrak D}_L$ coming from
the down sector. If, for the moment, we assume that ${\mathfrak D}_L$
is equal to the unit matrix to the accuracy desired, we can write the
corrected CKM matrix as
\begin{eqnarray}
V_{\rm CKM} =  {\mathfrak U}_L^\dagger {\cal U}_L^\dagger{\cal D}_L  \,.
\end{eqnarray}

We mentioned earlier that ${\mathfrak U}_L$ involves some small
parameter $\lambda$.  Following Wolfenstein \cite{Wolfenstein:1983yz},
we can use the Cabibbo angle for this parameter.  We define
\begin{eqnarray}
\lambda = \sin \theta_C \,,
\end{eqnarray}
and, motivated by the form for the correction matrix in \Eqn{Mupert},
write 
\begin{eqnarray}
{\mathfrak U}_L = \begin{pmatrix}
1 & A\lambda^2 & C\lambda^3 \\
A'\lambda^2 & 1 & 0 \\
C'\lambda^3 & 0 & 1 \\
\end{pmatrix} + \order(\lambda^4)\,,
\label{boldU}
\end{eqnarray}
which is consistent with our order of magnitude estimations in the
previous paragraphs.\footnote{Had we replaced $v_2$ through
  \Eqn{delta} in the expression for $\tilde{\cal M}_u$, the form for
  the correction matrix would have been different.  That would have
  obscured much of the subsequent discussions, although the final
  result should have been the same because physics should be
  independent of the parametrization.} We can choose the phases of the
quark fields such that the co-efficient $A$ is real in \Eqn{boldU}.
The unitarity of this matrix is achieved, to the same accuracy in
$\lambda$, by choosing
\begin{eqnarray}
A' = -A \,, \qquad C' = -C^* \,.
\label{AC}
\end{eqnarray}
In order to maintain consistency, we need to use the same accuracy of
$\lambda$ for the $\mathbb{Z}_2$ invariant CKM matrix given in
\Eqn{Z2CKM}.  Thus we obtain, correct up to terms of
$\order(\lambda^3)$, the following form for the CKM matrix:
\begin{eqnarray}
V_{\rm CKM} &=& \begin{pmatrix}
1 & -A\lambda^2 & -C\lambda^3 \\
A\lambda^2 & 1 & 0 \\
C^*\lambda^3 & 0 & 1 \\
\end{pmatrix} 
\begin{pmatrix}
1 & 0   & 0 \\ 0  & 1-\frac{\lambda^2}{2} & -\lambda \\ 0 & \lambda & 1-\frac{\lambda^2}{2} \\
\end{pmatrix}
 + \order(\lambda^4) \\ &=&
 \bordermatrix{\text{ } & \mathbf{b} & \mathbf{s} & \mathbf{d} \cr
                \mathbf{t} &1 &  -A\lambda^2   & A\lambda^3(1-\rho-i\eta) \cr
                                \mathbf{c} & A\lambda^2  & 1-\frac{\lambda^2}{2} & -\lambda \cr
                \mathbf{u} & A\lambda^3(\rho-i\eta) & \lambda & 1-\frac{\lambda^2}{2}
 } + \order(\lambda^4) \,,
\label{CKMW}
\end{eqnarray}
where in the last step, the CKM matrix in terms of the Wolfenstein
parameters are obtained by choosing
\begin{eqnarray}
C = (\rho+i\eta) A \,.
\end{eqnarray}

So far we have assumed that ${\mathfrak D}_L$ is the unit matrix.  This
is not an unjust assumption. The reason is that the mass hierarchy in
the down sector is not so violent as in the up-sector. So in the
expression for ${\cal M}_d^{\rm block}$ which is obtained by replacing
$y^u$'s with $y^d$'s in ${\cal M}_u^{\rm block}$, $y^d_1$ can carry
almost all the bottom quark mass while $y_2^d$ is very small.  Since
$y_4^d$ and $y_5^d$ only get involved into the expressions for strange
and down quark masses, they, too, are expected to be very small.
Therefore, the overall size of the perturbation matrix for the down
sector can be considered to be much smaller compared to that for the
up sector.  Hence at the leading order, ${\mathfrak D}_L$ can be
approximated as a $3\times 3$ unit matrix. But even if we consider the
small departures of ${\mathfrak D}_L$ from the unit matrix, we can use
a form like that for ${\mathfrak U}_L$ shown in \Eqn{Ucorr} with $A$
and $C$ replaced by different parameters. Then the CKM matrix will be
given by
\begin{eqnarray}
V_{\rm CKM} = {\mathfrak U}_L^\dagger {\cal U}_L^\dagger{\cal
  D}_L{\mathfrak D}_L \,.
\end{eqnarray}
However, this will not change the general form of the CKM matrix shown
in \Eqn{CKMW}. The Cabibbo block will not change at all to the order
shown and the Wolfenstein parameters, $A$, $\rho$ and $\eta$ will be
defined by linear combinations of parameters appearing in
${\mathfrak U}_L$ and ${\mathfrak D}_L$.

To conclude, we have shown that the quark masses and mixings can be
understood from a Lagrangian with an $S_3$ symmetry, broken
spontaneously down to $\mathbb{Z}_2$ by the VEVs that break gauge
symmetry. Regarding the masses, our analysis provides an explanation
of the third generation of quarks being widely different from the
first two --- the third generation is $\mathbb{Z}_2$-odd whereas the
first two are $\mathbb{Z}_2$-even.  Regarding mixing, we obtain the
Wolfenstein form of the CKM matrix. Wolfenstein fixed the orders of
$\lambda$ in various elements of the CKM matrix by experimental data
only.  In our case, we show that the remnant $\mathbb{Z}_2$ symmetry
ensures the form of the mixing matrix to $\order(\lambda)$.  Our
corrections to this order were motivated by consideration of a {\em
  heavy} top quark mass, and by terms which softly break the $S_3$
symmetry.  Without these soft breaking terms, the third generation
quarks do not mix at all with the other two generations.  Therefore,
smallness of $b \to s\gamma$ branching ratio can be related, in the
't\,Hooft sense of naturalness, to the smallness of the soft breaking
terms.

Tests of the idea presented here would consist of checking
consequences of the $\mathbb{Z}_2$ symmetry.  In the limit that the
$\mathbb{Z}_2$ symmetry is exact, in addition to the SM-like Higgs
$h$ which is $\mathbb{Z}_{2}$-even, there will be four neutral and
two pairs of charged spinless particles.  Among these, the scalar
$h^0$, the pseudoscalar $A_1$ and one pair of charged scalars
$H_{1}^\pm$ would be $\mathbb{Z}_2$-odd, and the others will be
$\mathbb Z_2$-even \cite{Das:2014fea}.  From the unitarity
considerations it has been shown \cite{Das:2014fea} that the masses of
these extra physical scalars are below 1 TeV.

Like most of the extended scalar sector models here also the FCNC
related issues are to be dealt carefully. Even though a dedicated FCNC
study of this model is beyond the ambit of this paper, we make a few
comments regarding this.  In the alignment limit the particle
$h$ will have exact SM-like couplings, and will not generate any
tree-level FCNC.  However, the other neutral scalars will in general 
have tree-level FCNCs, and their masses and couplings can be
constrained from flavor data.  For example, the $\mathbb{Z}_{2}$-even
scalars other than $h$ can induce FCNC involving the first two
generations of quarks, thereby contributing to  $K^{0}$-$\bar{K}^{0}$
oscillation.  Similarly, $\mathbb{Z}_{2}$-odd neutral scalars will be
constrained by neutral $B$-meson oscillation data.

Earlier, the signatures of $h^0$ were
studied~\cite{Bhattacharyya:2010hp, Bhattacharyya:2012ze}, but the
residual $\mathbb{Z}_{2}$ symmetry was not identified and thus the
generic behavior of other similar scalars was not realized properly.
For example, a light enough $h^{0}$ state can be probed in the $t\to
ch^{0}$ channel whereas a heavier $h^{0}$ can manifest itself in the
channel $h^{0}\to (t\bar{c}+c\bar{t})$. It is also worth mentioning
that in the exact $\mathbb{Z}_2$ limit, we do not have $\bar{t}th^0$
coupling and in the $S_3$ alignment limit~\cite{Das:2014fea}, $h^0VV$
($V=W,Z$) coupling also vanishes. But it might be possible to produce
$h^0$ via the coupling with the SM-like Higgs ($h^0h^0h$). Of course,
for testing any such outcome, it will have to be remembered that the
$\mathbb{Z}_2$ symmetry is violated by soft terms in the Lagrangian,
so the processes forbidden by the $\mathbb{Z}_2$ symmetry will
actually occur, although with a very small rate proportional to powers
of $\Delta/v$.  The decay of SM-like Higgs, $h\to \gamma \gamma$ can
also be useful in the sense that a precise measurement of this
diphoton signal strength can put constraints on the charged scalar
masses as they are not decoupled~\cite{Das:2014fea,
  Bhattacharyya:2014oka} even when their masses lie in the TeV range.

\paragraph*{Acknowledgements\,:}
We thank G.~Bhattacharyya for discussions. DD thanks the Department of
Atomic Energy (DAE), India for financial support.


\bibliographystyle{JHEP}
\bibliography{s3ckm-ref.bib}

\providecommand{\href}[2]{#2}\begingroup\raggedright\begin{thebibliography}{10}

\bibitem{Aad:2012tfa}
{\bf ATLAS} Collaboration, G.~Aad et~al., {\it {Observation of a new particle
  in the search for the Standard Model Higgs boson with the ATLAS detector at
  the LHC}},  {\em Phys.Lett.} {\bf B716} (2012) 1--29,
  [\href{http://arxiv.org/abs/1207.7214}{{\tt arXiv:1207.7214}}].

\bibitem{Chatrchyan:2012ufa}
{\bf CMS} Collaboration, S.~Chatrchyan et~al., {\it {Observation of a new boson
  at a mass of 125 GeV with the CMS experiment at the LHC}},  {\em Phys.Lett.}
  {\bf B716} (2012) 30--61, [\href{http://arxiv.org/abs/1207.7235}{{\tt
  arXiv:1207.7235}}].

\bibitem{Wolfenstein:1983yz}
L.~Wolfenstein, {\it {Parametrization of the Kobayashi-Maskawa Matrix}},  {\em
  Phys.Rev.Lett.} {\bf 51} (1983) 1945.

\bibitem{Mondragon:1999jt}
A.~Mondragon and E.~Rodriguez-Jauregui, {\it {The CP violating phase delta(13)
  and the quark mixing angles theta(13), theta(23) and theta(12) from flavor
  permutational symmetry breaking}},  {\em Phys. Rev.} {\bf D61} (2000) 113002,
  [\href{http://arxiv.org/abs/hep-ph/9906429}{{\tt hep-ph/9906429}}].

\bibitem{Kubo:2003iw}
J.~Kubo, A.~Mondragon, M.~Mondragon, and E.~Rodriguez-Jauregui, {\it {The
  Flavor symmetry}},  {\em Prog.Theor.Phys.} {\bf 109} (2003) 795--807,
  [\href{http://arxiv.org/abs/hep-ph/0302196}{{\tt hep-ph/0302196}}].

\bibitem{Chen:2004rr}
S.-L. Chen, M.~Frigerio, and E.~Ma, {\it {Large neutrino mixing and normal mass
  hierarchy: A Discrete understanding}},  {\em Phys.Rev.} {\bf D70} (2004)
  073008, [\href{http://arxiv.org/abs/hep-ph/0404084}{{\tt hep-ph/0404084}}].

\bibitem{Teshima:2005bk}
T.~Teshima, {\it {Flavor mass and mixing and S(3) symmetry: An S(3) invariant
  model reasonable to all}},  {\em Phys. Rev.} {\bf D73} (2006) 045019,
  [\href{http://arxiv.org/abs/hep-ph/0509094}{{\tt hep-ph/0509094}}].

\bibitem{Teshima:2012cg}
T.~Teshima, {\it {Higgs potential in $S_3$ invariant model for quark/lepton
  mass and mixing}},  {\em Phys.Rev.} {\bf D85} (2012) 105013,
  [\href{http://arxiv.org/abs/1202.4528}{{\tt arXiv:1202.4528}}].

\bibitem{Canales:2012ix}
F.~Gonzalez~Canales, A.~Mondragon, U.~S. Salazar, and L.~Velasco-Sevilla, {\it
  {$S_3$ as a unified family theory for quarks and leptons}},  {\em
  J.Phys.Conf.Ser.} {\bf 485} (2014) 012063,
  [\href{http://arxiv.org/abs/1210.0288}{{\tt arXiv:1210.0288}}].

\bibitem{Canales:2013cga}
F.~Gonz\'{a}lez~Canales, A.~Mondrag\'{o}n, M.~Mondrag\'{o}n, U.~J. Salda\~{n}a
  Salazar, and L.~Velasco-Sevilla, {\it {Quark sector of S3 models:
  classification and comparison with experimental data}},  {\em Phys.Rev.} {\bf
  D88} (2013) 096004, [\href{http://arxiv.org/abs/1304.6644}{{\tt
  arXiv:1304.6644}}].

\bibitem{Ma:2013zca}
E.~Ma and B.~Melic, {\it {Updated $S_{3}$ model of quarks}},  {\em Phys.Lett.}
  {\bf B725} (2013) 402--406, [\href{http://arxiv.org/abs/1303.6928}{{\tt
  arXiv:1303.6928}}].

\bibitem{Das:2014fea}
D.~Das and U.~K. Dey, {\it {Analysis of an extended scalar sector with $S_3$
  symmetry}},  {\em Phys.Rev.} {\bf D89} (2014) 095025,
  [\href{http://arxiv.org/abs/1404.2491}{{\tt arXiv:1404.2491}}].

\bibitem{Pakvasa:1977in}
S.~Pakvasa and H.~Sugawara, {\it {Discrete Symmetry and Cabibbo Angle}},  {\em
  Phys.Lett.} {\bf B73} (1978) 61.

\bibitem{Kubo:2004ps}
J.~Kubo, H.~Okada, and F.~Sakamaki, {\it {Higgs potential in minimal S(3)
  invariant extension of the standard model}},  {\em Phys.Rev.} {\bf D70}
  (2004) 036007, [\href{http://arxiv.org/abs/hep-ph/0402089}{{\tt
  hep-ph/0402089}}].

\bibitem{Koide:2005ep}
Y.~Koide, {\it {Permutation symmetry S(3) and VEV structure of flavor-triplet
  Higgs scalars}},  {\em Phys.Rev.} {\bf D73} (2006) 057901,
  [\href{http://arxiv.org/abs/hep-ph/0509214}{{\tt hep-ph/0509214}}].

\bibitem{Machado:2012ed}
A.~Machado and V.~Pleitez, {\it {A model with two inert scalar doublets}},
  \href{http://arxiv.org/abs/1205.0995}{{\tt arXiv:1205.0995}}.

\bibitem{Barradas-Guevara:2014yoa}
E.~Barradas-Guevara, O.~F\'{e}lix-Beltr\'{a}n, and
  E.~Rodr\'{i}guez-J\'{a}uregui, {\it {Trilinear self-couplings in an S(3)
  flavored Higgs model}},  {\em Phys. Rev.} {\bf D90} (2014), no.~9 095001,
  [\href{http://arxiv.org/abs/1402.2244}{{\tt arXiv:1402.2244}}].

\bibitem{Barradas-Guevara:2015rea}
E.~Barradas-Guevara, O.~F\'{e}lix-Beltr\'{a}n, and
  E.~Rodr\'{i}guez-J\'{a}uregui, {\it {CP breaking in $S(3)$ flavoured Higgs
  model}},  \href{http://arxiv.org/abs/1507.05180}{{\tt arXiv:1507.05180}}.

\bibitem{Bhattacharyya:2010hp}
G.~Bhattacharyya, P.~Leser, and H.~Pas, {\it {Exotic Higgs boson decay modes as
  a harbinger of $S\_3$ flavor symmetry}},  {\em Phys. Rev.} {\bf D83} (2011)
  011701, [\href{http://arxiv.org/abs/1006.5597}{{\tt arXiv:1006.5597}}].

\bibitem{Bhattacharyya:2012ze}
G.~Bhattacharyya, P.~Leser, and H.~Pas, {\it {Novel signatures of the Higgs
  sector from S3 flavor symmetry}},  {\em Phys.Rev.} {\bf D86} (2012) 036009,
  [\href{http://arxiv.org/abs/1206.4202}{{\tt arXiv:1206.4202}}].

\bibitem{Bhattacharyya:2014oka}
G.~Bhattacharyya and D.~Das, {\it {Nondecoupling of charged scalars in Higgs
  decay to two photons and symmetries of the scalar potential}},  {\em Phys.
  Rev.} {\bf D91} (2015), no.~1 015005,
  [\href{http://arxiv.org/abs/1408.6133}{{\tt arXiv:1408.6133}}].

\end{thebibliography}\endgroup
\end{document}